\documentclass{iau}
\usepackage{textgreek}
\usepackage{graphicx,natbib}
\newcommand{\apj}{ApJ}

\newcommand{\apjs}{ApJ}

\newcommand{\aap}{A\&A}
\newcommand{\aj}{AJ}

\title[Near-infrared abundances from high resolution spectra in the inner Bulge] 
{Detailed near-IR stellar abundances of red giants in the 
Inner Bulge and Galactic Center}

\author[N. Ryde, R. M. Rich, B. Thorsbro, et al.]   
{N. Ryde$^1$,
R. M. Rich$^2$, B. Thorsbro$^1$, M. Schultheis$^3$, T. K. Fritz$^4$,  \and L. Origlia$^5$}

\affiliation{$^1$Lund Observatory, Department of Astronomy and Theoretical Physics, \\ Lund University, Box 43, SE-221 00 Lund, Sweden \\ email: {\tt ryde@astro.lu.se} \\[\affilskip]
$^2$Department of Physics and Astronomy, UCLA, \\ 430 Portola Plaza, Box 951547, Los Angeles, CA 90095-1547, USA \\email: {\tt rmr@astro.ucla.edu} \\[\affilskip]
$^3$Observatoire de la C\^ote d'Azur, CNRS UMR 7293, \\ BP4229, Laboratoire Lagrange, F-06304 Nice Cedex 4, France \\email: {\tt mathias.schultheis@oca.eu} \\[\affilskip]
$^4$Department of Astronomy, University of Virginia, \\ 3530 McCormick Road, Charlottesville, VA 22904, USA
  \\email: {\tt tkf4w@virginia.edu} \\[\affilskip]
$^5$INAF - Osservatorio Astronomico di Bologna, Via Gobetti 93/3, \\ I-40129 Bologna, Italy \\email: {\tt livia.origlia@oabo.inaf.it}
}

\pubyear{2017}
\volume{334}  
\jname{Rediscovering our Galaxy}
\editors{C. Chiappini, I. Minchev, E. Starkenburg, \& M. Valentini., eds.}
\begin{document}

\maketitle

\begin{abstract}

Owing to their extreme crowding and high and variable extinction, stars in the Galactic Bulge, within $\pm2^\circ$ of the Galactic plane, and especially those in the Nuclear Star Cluster, have only rarely been targeted for an analyses of their detailed abundances. There is also some disagreement about the high end of the abundance scale for these stars. It is now possible to obtain high dispersion, high S/N spectra in the infrared K band ($\sim2.0-2.4\,$\textmu m) for these giants; we report our progress at Keck and VLT in using these spectra to infer the composition of this stellar population.

\end{abstract}

\firstsection 
\section{Introduction}
High resolution spectroscopy has been employed to infer the chemistry and history of populations across the Milky Way. Advances in infrared detector technology and instrumentation have recently enabled large scale studies directed toward the central parts of the Galaxy - the bulge and nuclear star cluster- where extinction and crowding are both high.   We refer in particular to the inner $\pm2^\circ$ of the Galactic plane that has escaped many detailed investigations due to extreme optical extinction caused by dust in the line of sight.  It is easy to forget that for some of the red giants in the central cluster, SgrA* is closer than Proxima Centauri is to the Sun.  How does such a unique environment affect chemical evolution?

\begin{figure}[!hbt]
\begin{center}
 \includegraphics[angle=-90,width=5.5in]{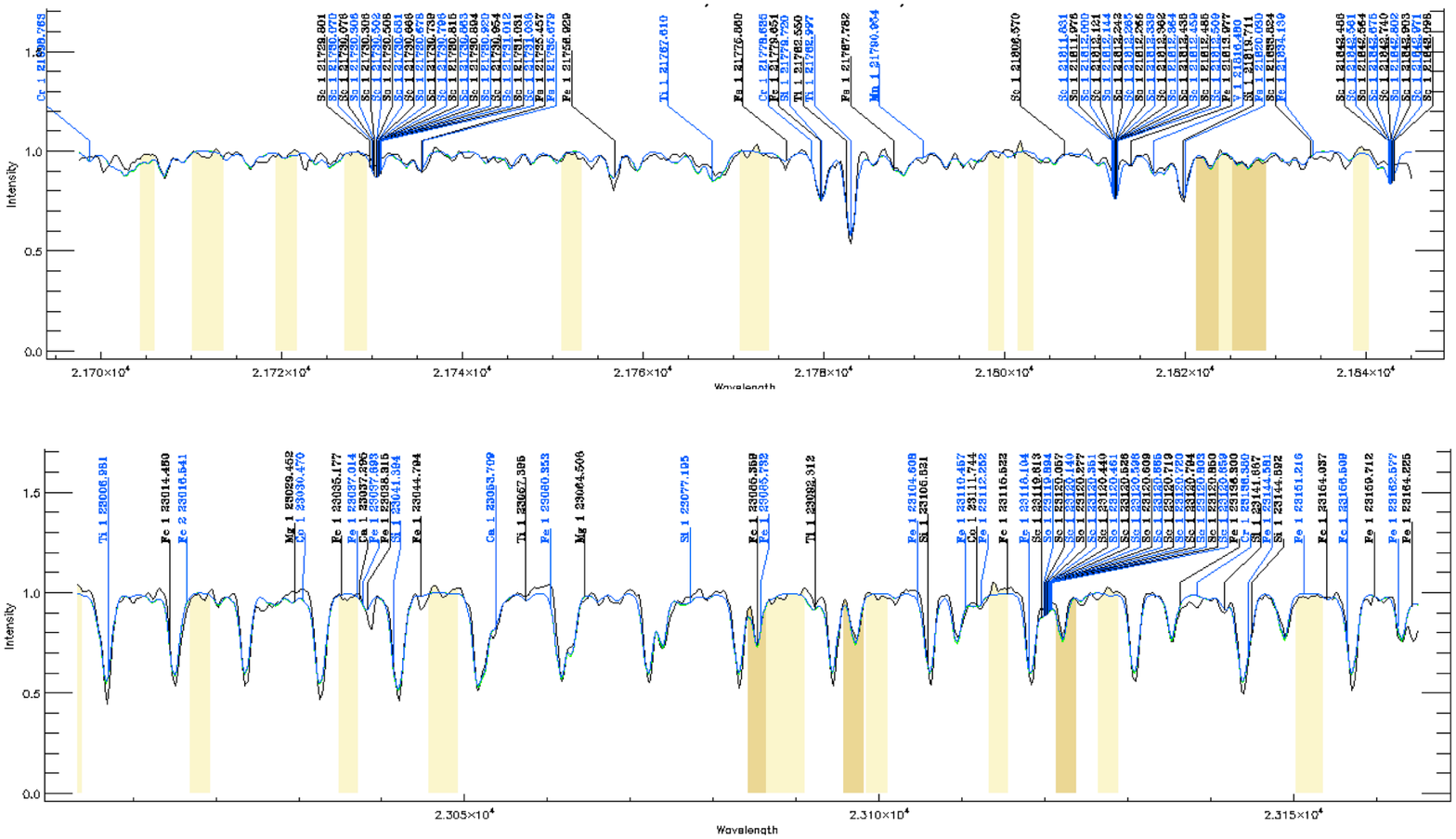} 
 \caption{Example of a spectrum of a Galactic Center giant observed with KECK/NIRSPEC at a spectral resolution of $R=23\,000$ for an integration time of 40 minutes, achieving SNR = 100. Some atomic lines are marked. The brown bands show the regions with good CN and CO lines for the simultaneous fit of the C and N abundances. The yellow bands show good continuum bands for the normalization of the spectra. This M giant ($K_S = 11.0$) has determined stellar parameters of $T_\mathrm{eff}=3900\,\mathrm{K}$, $\log(g)=1.5$, and [Fe/H]$=+0.1$ (for details see Rich et al. 2018.). 
}
   \label{fig1}
\end{center}
\end{figure}

We are engaged in a long-term project employing high resolution infrared spectroscopy to lift the reddening veil  and to explore the Galactic bulge and center in the near-IR. For an accurate abundance determination, we need high resolution, high S/N spectra of red giants, which we can achieve using K-band spectroscopy observed at KECK with the NIRSPEC spectrometer \citep{nirspec} and the VLT with the CRIRES spectrometer \citep{crires1}. We also need a line list at least as good in the near-IR as is available in the optical.  It is now possible to derive abundances for the light elements as well as iron, for red giants in the Nuclear Star Cluster at the center of the Galaxy \citep[see, e.g. Figure \ref{fig1} and][]{ryde:16}. 

\section{The stellar metallicity distribution and $\alpha$-element trends in the Bulge}

The metallicity distribution of the stars in the bulge are shown to be wide, from approximately [Fe/H]$=-1.5$ to $+0.5$ or more \citep[see e.g.][]{johnson:13,schultheis:15,rojas:17}. The question of the bimodality of this distribution is still debated; although gaussian mixture models applied to the abundance distribution support bimodality, corollary support for bimodality from other physical parameters e.g. kinematics and composition is not robust \citep[e.g.][]{johnson:14,rojas:17}.  The composition trends appear similar to that of the thick disk, but significant trends differ \citep[]{johnson:14} and of course, the thick disk never reaches [Fe/H]=+0.5 as seen in the bulge. A vertical metallicity gradient, with increasing metallicity with decreasing Galactic latitude is generally found down to approximately 500 pc. At distances of 140-400 pc from the nucleus, \citet{rich:12} find no abundance gradient with $\langle \rm [Fe/H] \rangle=-0.2$. The alpha-element trends are, however, found to be similar in all fields all the way into the center \citep{johnson:11,ryde:16}. 

The Nuclear Star Cluster, is of special interest as it is projected within the sphere of influence of Sgr A$^*$, surrounded by the Bulge. While the inner bulge is predominantly old, the Cluster also has ongoing star-formation. It can therefore not be assumed that these populations should be of the same origin. There are several formation routes which can be tested by determining the metallicities of the stars in the Cluster. One of our goals is to connect old stars in Galactic Center and in the Bulge, which can be done with a homogeneous data set, using same analysis technique.

\section{Near-IR Spectroscopy - Challenges}

Since stars in the inner Bulge and in the Nuclear Cluster are faint and cool, the spectra are difficult to analyze. The cooler the star, the greater the impact of molecular absorption on the entirety of the spectrum and the risk of unaccounted blending of molecular and atomic lines, even at high dispersion. Further, the K-band  is a relatively new wavelength region being used for stellar abundance analysis, meaning that work has to be put into developing the methods to analyze the spectra; a final challenge is that many lines of the light elements of interest are present but too strong for abundance analysis. A careful analysis is required and the observations have to be optimized to minimize systematics. High spectral resolution is needed for accurately determining the composition of these bulge giants. 

Elements that have spectral lines appropriate for an abundance analysis are Fe, C, N, F, Sc, Na, Al, and the alpha elements Mg, Si, S, Ca, and Ti.  A new atomic line list is needed (Thorsbro et al., these proceedings) but the important CN lines are well described with the list provided in \citet{sneden:14}. 

In the near-IR, spectral lines saturate at a smaller line depth than at shorter wavelengths. Strong lines should therefore be avoided since they get increasingly abundance insensitive, and are increasingly sensitive to the largely unknown microturbulence parameter. Furthermore, the cores can be sensitive to non-LTE effects, such as scattering, and they might probe the outer layers of the atmosphere where the physics is uncertain. In low-resolution spectra, and especially at low signal-to-noise levels, one might be tempted to use these strong lines.

\section{Results}

\begin{figure}[!bht]
\begin{center}
 \includegraphics[angle=90,width=6.3in]{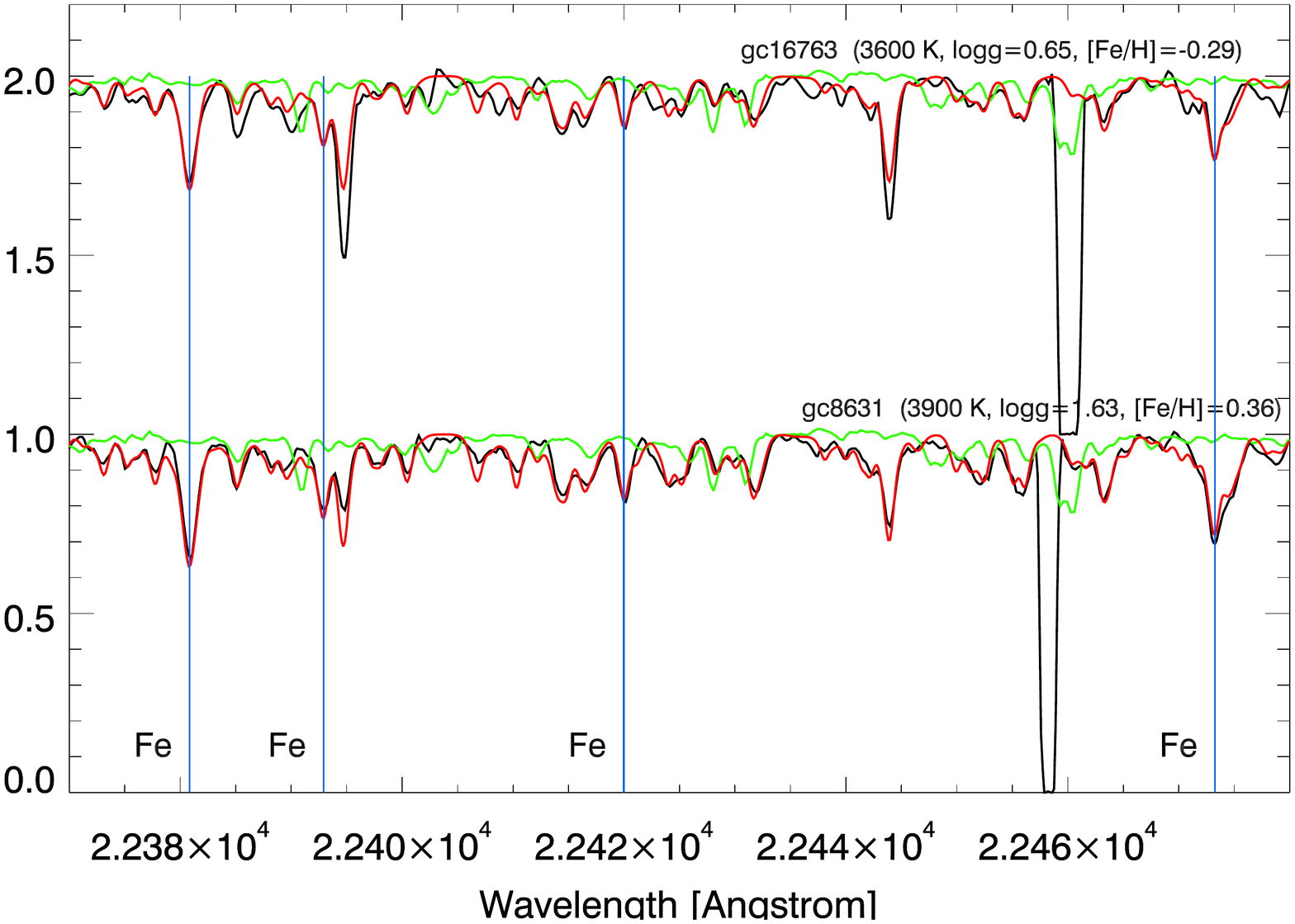}
 \caption{Here we present analysis of two of the observed Galactic Center stars, one metal-poor
and one metal-rich, shown with the black spectra. The red line is the synthetic spectra. The green spectra show the telluric features. The Fe lines in this wavelength range used for the metallicity determination, are marked with blue vertical lines.
We use effective temperatures determined by low resolution observations of the CO-bandhead
at $2.3\,\mu$m and $\log(g)$ values are based on de-reddened photometry (for details see Rich et al. 2018, in prep.). 
}
   \label{fig2}
\end{center}
\end{figure}

After analysing VLT/CRIRES spectra of fields at latitudes $b=0$, $-1$, and $-2^\circ$ in \citet{ryde_schultheis:15,ryde:16}, a total of $\simeq 50$ stars in the corresponding fields at Northern latitudes are under analysis,  Ca.\ 25 giants in the Nuclear Star Cluster have been observed with Keck/NIRSPEC \citep[][Rich et al.2018, and Thorsbro et al. 2018]{ryde:16}. Figure \ref{fig2} shows stellar spectra of two such giants in the Nuclear Star Cluster. We see a broad metallicity distribution with few stars having [Fe/H]$<-1$ (less than $\sim5\%$) and we do not detect extremely metal-rich stars. A full metallicity distribution will be presented in Rich et al. (2018, in prep.) and the $\alpha$-element trends in Thorsbro et al. (2018 in prep.).

\bibliographystyle{aa}

\end{document}